# The Three Component Electronic Structure of the Cuprates Derived from SI-STM

J.W. Alldredge[1], K. Fujita[2], H. Eisaki[3], S. Uchida[4], Kyle McElroy[1]

1 University of Colorado Boulder, Boulder CO 80309

2 LASSP, Department of Physics, Cornell University, Ithaca NY 14853

3 Institute of Advanced Industrial Science and Technology, Tsukuba, Ibaraki 305-8568, Japan

4 Department of Physics, University of Tokyo, Bunkyo-ku, Tokyo 113-0033, Japan

**Abstract:**

We present a phenomenological model that describes the low energy electronic structure of the cuprate high temperature superconductor $Bi_2Sr_2CaCu_2O_{8+x}$ as observed by Spectroscopic Imagining Scanning Tunneling Microscopy (SI-STM). Our model is based on observations from Quasiparticle Interference (QPI) measurements and Local Density of States (LDOS) measurements that span a range of hole densities from critical doping, p~0.19, to extremely underdoped, p~0.06. The model presented below unifies the spectral density of states observed in QPI studies with that of the LDOS. In unifying these two separate measurements, we find that the previously reported phenomena, the Bogoliubov QPI termination, the checkerboard conductance modulations, and the pseudogap are associated with unique energy scales that have features present in both the ***q***-space and LDOS(E) data sets.

**Intro:**

One of the challenges to understanding SI-STM results is interpreting the large amount of data each measurement produces. Today, a typical mapping of the energy resolved differential conductance, in a 50nm$^2$ field of view, can easily produce 10$^7$ data points within ten days. Such datasets contain a wide variety of features and phenomena, both in the energy resolved local density of state (LDOS(E)) curves and in the spatial modulations of the density of states (LDOS(***r***)). When the LDOS(r) is Fourier transformed (FT), the coherent spatial modulations are represented as peaks in ***q***-space. In order for the data to be interpreted and compared to theory, as well as other experimental probes, it must be reduced to its key observables. In the past, individual data sets were analyzed through either mapping out the QPI ***k***-space scatters via the octet model[1–9], or by looking at certain features/spatial patterns in the

LDOS[10–13]. This has left information present in the data sets unreported and unanalyzed. In order to process this data, a method for parameterizing the LDOS($r$,E) must be developed. A successful method will need to be able to parameterize a wide range of noisy LDOS(E) data, accomplish this parameterization consistently with the parameterization of the *q*-space data, and provide new information about the relationship between *q*-space phenomena and the structure present in the individual LDOS(E) curves. The method presented here, is based on a model spectral density of states (A(*k*,E)). This *k*-space density of states can be matched to the QPI derived data in *q*-space using the octet model, and by integrating the A(*k*,E) over *k*-space, it can be matched to the LDOS(E) data.

The model presented here for the spectral density of states consists of three different regions in energy. The spectral density of states that is generated by this tripartite model is a *k*-space structure that reproduces the *q*-space and individual LDOS(E) data. This *k*-space A(*k*,E) is not unique in producing a spectral density of states that matches the data, and we are unable to claim that the A(*k*,E) represents the underlying physics, only that it produces the spectral density of states that is consistent with both the QPI data and the LDOS(E) data. Despite this limitation, the spectral density of states generated with the tripartite model has key energy scales that are associated with transitions in its *k*-space structure. The energy and momentum at which these transitions take place represent simultaneous transitions in *q*-space and the LDOS(E).

We use five dI/dV($r$,E) data sets that consist of five different dopings of $Bi_2Sr_2CaCu_2O_{8+x}$ to test the model. This allows us to validate the model for a wide range of dopings in terms of the average LDOS(E) and the QPI data. All of these data sets have had particular features in them analyzed and reported on in the past[1,4,10–12,14] with summaries of the phenomenological overview being provided more recently[5].

In this study, we present our model, test its validity in *k*-space and LDOS(E) using the 5 data sets, tie the energies defined by the model to *q*-space and LDOS(E) phenomena and discuss how the spectral density of states inferred here relates to other probes measures of the electronic dispersion.

**Background:**

Previously[10], the LDOS(E) has been describe using a d-wave Dynes style[15] model. That study focused on the scattering present in the LDOS(E). The Dynes model used previously is described by the following equation:

$$N(E,\Gamma_2) = A \times \left| \text{Re}\left( \left\langle \frac{E + i(\Gamma_1 + \Gamma_2(E))}{\sqrt{(E + i(\Gamma_1 + \Gamma_2(E)))^2 - \Delta(k)^2}} \right\rangle_{fs} \right) \right| + C \times E \qquad (1)$$

where A is a prefactor used to scale the model LDOS(E) to match the data, E is the energy, $\Gamma_1$ is the unitary scattering rate, $\Gamma_2$ is the linear in energy high energy scattering rate, $\Delta(k)$ is a d-wave gap, represented by Cos[2*θ], and C is a linear background term needed to match the data's background. In order to calculate a LDOS(E) curve, the equation is integrated along the Fermi surface. It is important to note that $\Delta(k)$ was not necessarily proposed to be the gap of a simple d-wave superconductor, although to first order it is[10].

This Dynes model was chosen previously, with its addition of an energy dependent lifetime ($\Gamma_2$), in order to fit a large section of the LDOS(E) measurements over several dopings using the fewest number of free parameters. The two different lifetime terms are necessary in order to match the two scattering phenomena observed in the LDOS[10]. Around unitary scatters[16], such as vacancies and zinc impurities, there is a suppression of the gap peak, as well as an increase in the number of low energy states[17]. Away from unitary scatters, the spectra display a suppressed d-wave gap peak. These smaller peaks were modeled with a scattering term that increases linearly in energy ($\Gamma_2(|E|)=\alpha*|E|$). This second scattering term draws its inspiration from Born scattering[16], although it is more likely caused by impurity-plus-spin fluctuations[18]. $\alpha$ is a necessary parameter to fit the data if thermal broadening or background noises do not dominate the measurement.

Despite the success of the Dynes model, it was observed in the previous study[10] that the model failed to reproduce the LDOS(E) at low dopings/high gap values. Figure 1 shows representative spectra taken from five different dopings showing good agreement with the d-wave model (the dotted black line), until the doping becomes low ($p<0.10$). There exists a 'kink' in the LDOS(E) at low dopings that is unaccounted for in the previous studies. The energy span of the kink increases with decreasing doping, causing the Dynes model to fail at low dopings.

The energy dispersion extracted from the QPI[1,4,7,8] shows a clear departure from a lowest order d-wave dispersion. The QPI dispersion cannot be described by a simple d-wave gap and superconducting A(k,E). However the addition of a higher order d-wave term[2,4,19,20] allows the dispersion to be successfully matched. The addition of this higher harmonic gap term is in contrast to the previous LDOS(E) study[10] that relies on the lowest order d-wave gap. This difference between the two measurements leads to two separate descriptions of the electronic structure, one based on *q*-space measurements, the other based on LDOS(E).

QPI measurements do not give the underlying **k**-space electronic dispersion directly. The **q**-space pattern measured by the SI-STM instead measures scattering vectors between points of high **k**-space density of states. In order work backwards from the **q**-space to the **k**-space structure, the octet model[9] is employed. The octet model uses Fermi's golden rule, a superconducting dispersion with a doped Mott insulator band structure, and a d-wave gap, to determine that there are 8 points in **k**-space that will have large spectral density of states and will be the source and the destination for scattering **q**-vectors. These 8 points define 7 **q**-vectors connecting them and by measuring these **q**-vectors, one can work back to the original 8 points. By determining the 8 **k**-points as a function of energy, the dispersion and the gap can be measured. This model has been highly successful in determining the **k**-space structure of the electronic states and its results have been shown to agree with the joint density of states models when probe effects were taken into account[7,21]. In figure 2b the previously reported[4] **k**-space QPI points with their d-wave + higher harmonic fits are shown. The dotted lines in figure 2b represent the lowest order d-wave dispersions and are a clear departure from the data. The gap used to fit the QPI data is[2,19,20]:

$$\Delta_{QPI}(\theta) = \Delta_1 \cdot \left[ B \cdot Cos\ 2\theta\ +\ (1 - B) \cdot Cos\ 6\theta \right] \tag{2}$$

where B is allowed to vary from 1 to 0.7. There have been reports that the presence of the higher harmonic term is due to the incompleteness of the model used to extract the origin of the QPI peaks[22], which contradicts other studies[7,21]. The origin of Cos[6θ] term has also been attributed to a two particle scattering or a tunneling matrix element effect, both of which would be detectable by the SI-STM, and not necessarily detectable by other probes[23].

While SI-STM is able to measure a LDOS(E) across a wide range of energies (relevant to this study are energies below 150 meV), the dispersive QPI pattern terminates, or the signal to noise ratio of the individual **q**-space peaks becomes too low to resolve, at an energy around 35 meV. This termination energy occurs as the QPI determined dispersion reaches the region of **k**-space distinguished by lines connecting $(0,+-\pi/a_0)$ to $(+-\pi/a_0,0)$. These lines represent the anti-ferromagnetic zone boundary as it exists in the parent compound[4] and are shown as the dashed diagonal line in figure 2c. Here the QPI **k**-space scattering origins for the five samples used in this study[4] are plotted. The termination points are marked as the large green circles and represent the highest **k**-space momentum determined by the octet model. These momentums correspond to the highest energy where SI-STM can measure enough QPI peaks to use the octet model to determine the **k**-space momentum.

There are six main SI-STM observed phenomena present in the LDOS(**r**,E), three are present in single LDOS(E)'s and three are present as coherent spatial excitations and show up clearly in **q**-space. In figure 3, an overview of each phenomenon as well as roughly their ordering in energy is provided. The coherent QPI and the low energy V-shape in the LDOS are shown at low energies[1,11]; the kink and the checkerboard pattern at $q_1^* \sim \pi/2a_0$ follows in energy, existing between the two dotted lines[2,6,10]; the pseudogap peak and its spatial modulations pattern exits at high energies[14,24–26]. On a gross scale, the pseudogap spatial modulation consists of a modulation at $q_5^* \sim 3\pi/2a_0$. However, it has a more complicated finer structure[24–26]. All of these phenomena have been previously reported on, but no quantitative energies have been defined as boundaries between the regions.

The low energy V-shaped gap in the LDOS(E) is seen throughout the phase diagram and has been shown to exists at low energies where the coherent QPI pattern is also observed[5]. This coherent Cooper pairing phenomena[27] appears to end near the lines connecting $(+-\pi/a_0, 0)$ to $(0, +- \pi/a_0)$ where the QPI intensity fades out. At similar energies to the termination, we see a kink in the LDOS(E) that takes the form of a plateau. The kink phenomenon represents a departure from the d-wave Dynes model. The checkerboard phenomena or pattern has been reported in the past as a $q_1^* \sim \pi/2a_0$ wavelength pattern that exists on the underdoped side of the phase diagram and is present only in large gaped regions of the field of view[2]. In order to discuss the checkerboard modulations, we must first define exactly the phenomena we mean due to different definitions and observations[2–6,24,28–30] used in previous reports. This is especially important since the checkerboard modulations are a long wavelength phenomenon and there exists other phenomena that share similar long wavelengths (gap disorder, kink disorder, surface disorder) making the accurate identification of the checkerboards energy range difficult (see figure 7 for definition).

At higher energies in the LDOS(E), a spatially disordered pseudogap peak structure is observed[11]. These disordered peaks have been analyzed in the form of a map of the energy at which they occur[11,13], and this gap map disorder has been shown to be independent of doping[10] from p~0.19 to p~0.08. Coincident with this disordered gap energy, there exist disordered states that break the expected 90°-rotational ($C_4$) symmetry of electronic structure within $CuO_2$ unit cells, at least down to 180°-rotational ($C_2$) symmetry (nematic), but in a spatially disordered fashion. This intra-unit-cell $C_4$ symmetry breaking coexists with incommensurate conductance modulations locally breaking both rotational and translational symmetries (smectic). The wavelength of these modulations is also determined, empirically, by the **k**-space points where Bogoliubov quasiparticle interference terminates, and therefore evolves continuously with doping[5,14,25,26].

The primary reason behind the development of a new model put forward here, is to reconcile the two different measurements, one in the individual LDOS(E) and one in the spatial **q**-space excitations, of the electronic structure measured by SI-STM. We also aim understand the QPI termination at the parent compounds AF-zone boundary (the line $(+-\pi/a_0, 0)$ to $(0, +- \pi/a_0)$) and the kink in the LDOS(E), as spectral density of state features. By having a holistic description of the low energy electronic structure measured by SI-STM, we can reconcile differences between this local probe and global measurements such as Angular Resolved Photo-Emission Spectroscopy (ARPES), Raman spectroscopy, and thermal transport. A model that includes the kink, and is constant with the *q*-space QPI pattern, will also be important for future local studies of the variation in the LDOS(*r*,E).

**Model:**

The combination of the previous LDOS(E) d-wave gap and lifetime terms, and the observed QPI d-wave + higher harmonic gap is the central synthesis of the tripartite model. The low energy coherent QPI states have a gap structure that differs from the lowest harmonic term that was used to describe the LDOS(E) in previous fitting analysis[10]. The combination of the two gap structures results in a tripartite electronic structure that is illustrated in figure 4. At low energies/momentum, there is an effective gap of the form:

$$\Delta_{low}(\theta) = \Delta_1 \cdot [B \cdot Cos\ 2\theta\ +\ (1-B) \cdot Cos\ 6\theta] \qquad (3)$$

taken from the QPI analysis. This QPI derived section of the gap is the red curve in figure 4. The high energy/momentum gap used is a lowest order d-wave gap that was successful in past fitting analyses of the LDOS(*r*,E), and takes the form here of:

$$\Delta_{high}(\theta) = \Delta_1 \cdot Cos\ 2\theta \qquad (4)$$

this LDOS derived gap is shown in blue in figure 4. Even though these two effective gap structures have the same anti-nodal energy, there exists a discontinuity between the two. This discontinuity defines two additional energy values, the last energy where the higher harmonic term exists, $\Delta_0$, and the first energy where the lowest harmonic d-wave effective gap exists, $\Delta_{00}$. In *k*-space the angle of the cross over between the two gap structures ($\theta_{cross}$) is defined by

$$\Delta_0 = \Delta_1 \cdot [B \cdot Cos\ 2\theta_{cross}\ +\ (1-B) \cdot Cos\ 6\theta_{cross}] \qquad (5)$$

$$\Delta_{00} = \Delta_1 \cdot Cos\ 2\theta_{cross} \qquad (6)$$

The region between these two energy values is represented in our model by an energetically non-dispersive region, the simplest possible way to link the two gap structures. However, this region must have a small dispersion in order not to double value the greens function. The amount of dispersion that exists in this region is set by the ability of the SI-STM to resolve the checkerboard wavelength. We will see that the sharpness of this dispersion is set by the checkerboards non-dispersing signature which is ~$0.2\pi/a_0$ in width and has been measured[5] to be non-dispersive within ~$0.03\pi/a_0$.

The two part gap defined here for the model, is not necessarily the underlying physical gap of the system. The gap has been defined using the previous models that have been implemented for QPI and the LDOS(E) studies. The combination of the two model gaps used previously allows the creation of a spectral density of states that matches the data. However there are a number of methods that one might generate such a spectral density of states. Instead of focusing on the gap structure, we focus on the generated spectral density of states and the two new distinct energies $\Delta_0$ and $\Delta_{00}$ it contains. These energies represent distinct transitions in the spectral density of states that capture features in the LDOS and the QPI data.

The tripartite model defines three regions: 1) a low energy d-wave + higher harmonic, 2) an energetically non-dispersive transition region and 3), a high energy lowest order d-wave gap. The flexibility in the model allows the energy transition between the low energy d-wave + higher harmonic and the high energy lowest order d-wave to change freely. From the previous QPI termination studies[4], this transition will happen at the lines connecting (+-$\pi/a_0$,0) to (0,+-$\pi/a_0$) with a $\theta_{cross}$ set by the intersection of the line and the Fermi surface. However, this transition point in **k**-space is unconstrained by the model and is treated as an independent variable for the purposes of the fit. By fitting over the 5 dopings, the termination point can be determined and compared to the (+-$\pi/a_0$,0) to (0,+- $\pi/a_0$) lines.

The superconducting spectral density of states[31], A(**k**,E), is combined with unitary scattering lifetime and a linear in energy lifetime terms that were used previously with the Dynes model. The spectral density of states takes the form:

$$A(\vec{k},E) = \frac{-1}{\pi} \frac{\mathrm{Im}\Sigma(\vec{k},E)}{(E-\varepsilon(\vec{k})-\mathrm{Re}\Sigma(\vec{k},E))^2 + \mathrm{Im}\Sigma(\vec{k},E)^2} \qquad (7)$$

With the self-energy[29] described by

$$\Sigma(\vec{k},E) = -i\Gamma + \frac{\Delta_k^2}{E+\varepsilon(\vec{k})+i\Gamma} \qquad (8)$$

where $\varepsilon(\mathbf{k})$ is the tight binding band structure[32] that has had its chemical potential fit to the QPI data. The lifetime terms are defined as:

$$\Gamma = \Gamma_1 + \Gamma_2(E) \qquad (9)$$

$$\Gamma_2(E) = \alpha \cdot |E| \qquad (10)$$

The gap structure is the combination of the QPI gap and the LDOS(E) gap,

$$\Delta\theta = \begin{matrix} \Delta_1 \cdot Cos\ 2\theta_{cross} & for\ \theta > \theta_{cross} \\ \Delta_1 \cdot\ B \cdot Cos\ 2\theta_{cross}\ +\ 1-B\ \cdot Cos\ 6\theta_{cross} & for\ \theta < \theta_{cross} \end{matrix} \qquad (11)$$

For the dopings presented here, there is a small, in comparison with the $\Delta_1$ gap peak, Van Hove singularity on the positive side, which is suppressed by the energy dependent lifetime term. By integrating the A($k$,E) over the Brillion zone a LDOS(E) curve (figure 4d) is generated and by looking at the maximum density of states in $k$-space the octet model extracted $k$-space gaps (figure 4c) can be matched to the A($k$,E). For the LDOS(E) an additional linear background term is added to match the varying background present in the data. The QPI and LDOS(E) data represent two separate measurements of the spectral density of states, although both limited in different ways. The QPI provides k-space data up to the (+-π/$a_0$,0) to (0,+- π/$a_0$) boundary lines. The LDOS(E) provides energy data, but no $k$-space data. The tripartites model can generate $k$-space and LDOS(E) data, and allows both data sets to be fitted with one model.

**Results and Inferences:**

### Model fitting:

The test of the proposed tripartite model is whether or not it can adequately reproduce the data with a unique parameterization that can be tied to other observables. In figure 5 the fits for both the LDOS(E) and the $k$-space QPI origins show that the tripartite model has the ability to match both QPI and LDOS(E) data when each data set is fitted separately. The non-dispersive region in the model, between $\Delta_0$-$\Delta_{00}$ in energy and at $\theta_{cross}$, shows up as a kink structure when the LDOS is calculated by integrating the A($k$,E) over the Brillion zone. $\Delta_0$ marks the beginning of the kink, and $\Delta_{00}$ the end. This region is marked in green in figure 5. The tripartite model is an improvement over the previously used Dynes based model for the LDOS(E) fitting, since it is able to successfully track the kink structure as it grows larger with deceasing doping.

The kink starts as a tiny departure from the d-wave background, easily observable near p~0.14 and evolves to a large plateau at p~0.06. At higher dopings the kink blends into the $\Delta_1$ and is picked up by the fitting, although it is usually spaced close to the energy resolution of the data (2meV). Hence, there is some question about the presence of the kink at dopings near optimal. While the kink disappears for the average spectra at these high doping, for all the dopings up to and including p~0.19 (the highest analyzed in this study) there are always some spectra in a 50nm$^2$ field of view that contain the kink.

Figure 5 shows that the tripartite model reproduces both *q*-space and LDOS(E) data separately; however, in order for our model to be valid, the parameterization obtained for both the LDOS and the QPI must be the same, or nearly so if our model is to accurately represent the electronic structure of the cuprates. In Figure 6, the data and fits for both LDOS (right) and QPI (left) utilizing the same set of parameters for both are plotted. Procedurally each data set was fitted separately and then the resulting parameters were averaged. This results in good agreement across all dopings. The kink begins at the energy where the QPI terminates and is caused, in the model, by the non-dispersive section of the spectral density of states. The average peak determined gap values of the LDOS(E) are shown as open blue circles on the QPI side. While the higher energy dispersion, past ~40-60 meV depending on doping, in *q*-space cannot be measured, the fits to the QPI produce the same anti-nodal gap energy as the measured peak value in the LDOS(E). All of the average fitting parameters are listed in table 1 for ease of comparison.

The gap used in the tripartite model follows from the historical analysis used for both data sets. Each of the energy scales present in the model is indicative of a feature in the LDOS and QPI and each has other observables that agree with the energy values found. These other observables give additional meaning to each of the three energy scales, $\Delta_0$, $\Delta_{00}$, and $\Delta_1$, as well as verify that the tripartite spectral density of states is correct, at least in the momentum range where SI-STM can observe *q*-vectors (up till the (+-π/a$_0$,0) to (0,+- π/a$_0$) line). In order to identify these other observables the spatial excitations that are present above the QPI in energy need to be understood.

**Tripartite *q*-space excitations:**

The electronic spatial excitations display three separate identifiable regions. At low energies the dispersive octet QPI pattern exists; above that the checkerboard $q_1^*$ pattern is found and at higher energies still, the pseudogap $q_5^*$ pattern is present. The QPI pattern is defined as a dispersive pattern of an octet of *q*-space peaks and has been reported on in the past[1,3–5,8,9]. An example of the pattern is shown in figure 2a. Line cuts in *q*-space along the atomic direction, in figure 7, illustrate the signature of both the checkerboard and the pseudogap, as well the differences between the raw FT (a) and the ratio map FT (b). Figure 7 uses the data from a UD45k sample[24] as an example. This doping is low enough that the majority of the field of view displays both the checkerboard and the pseudogap. The FT of the data and ratio map have had each energy normalized in intensity, by setting the maximum intensity to 1, in order to accentuate the fine structure, and eliminate overall LDOS scaling changes. This is equivalent to the normalization process used when a movie or series of 2D individual energy QPI patterns are shown.

The negative energy LDOS(E) has features that depend on the setup conditions that the SI-STM uses. The setup effect comes from the SI-STM keeping the tunneling current constant at each spatial point, which modulates the tips position based off the integral of the positive energy LDOS's. For the LDOS analysis, the negative energies are not fit due to the presence of a setup effect that modulates the background and peak intensity as a function of setup current and is beyond the capabilities of our model to understand. This setup effect becomes more pronounced as the doping is decreased and could be due to the tip-sample separation decreasing as the total density of states decreases. When discussing the spatial excitations, these setup effects become important to determining actual spatial modulations apart from modulations caused by the SI-STM setup effect. The ratio of the positive to negative LDOS($r$,E) is used to remove some of these effects and it is why the negative energies are discussed here, and not elsewhere in the paper.

In the raw FT at low wave-vectors is $q_1$, which is marked by dotted black lines in figure 7a. On the positive side, between $\Delta_0$ and $\Delta_{00}$, there is also a non-dispersive feature, $q_1^*$, which overlaps the $q_1$ vector but is separate from it. This $q_1^*$, as labeled here, is the checkerboard signature, a non-dispersive coherent modulation that is separate from the $q_1$ dispersion. The $q_1^*$ defines the checkerboard as a separate modulation from the $q_1$ dispersion. While $q_1$ appears to disperse to lower wave lengths at higher energy values there is, circled in black, a non-dispersive structure almost on top of $q_1$. This $q_1^*$ is not present in the negative energy data; however, we see higher energy streaking at this wave vector on the negative side that is likely due to a setup effect.

$q_5$ appears around ~0.7 $2\pi/a_0$, disperses below $\Delta_0$ and then disappears at energies above $\Delta_0$. Above $\Delta_{00}$, there is the appearance of the static $q_5^*$ feature that exists at the same wave length as the $q_5$ peak at $\Delta_0$. This $q_5^*$ is the pseudogap signature or Electric Cluster Glass (ECG) reported previously[24], although here, it is seen in the raw data as opposed to the ratio map. Streaking appears on the negative side of the data due to the setup effect at this wavelength.

All recent QPI analysis has concentrated on the ratio map, which is the positive energy data divided by the negative energy data, rather than the raw data[5,24]. Looking at the same atomic direction cut as in 7a, 7b shows the FT ratio map line cut. There are a number of differences in the ratio map atomic direction line cut, firstly, $q_5$ disappears; this is thought to be caused by $q_5$ having the same phase between positive and negative energies[33]. Secondly, $q_1$ in the ratio map shifts in position and the extended $q_1$ that disperses to high energies and low wavelengths disappears. Thirdly, the checkerboard pattern is enhanced in strength and merges with the shifted $q_1$. The shift in $q_1$ is likely caused by a wavelength difference in the positive and negative dispersions and therefore the ratio map isolates the overlap between the two dispersions. This shift and change in $q_1$ could be the cause of some of the confusion in differentiating between the low energies coherent QPI $q_1$ and the checkerboard pattern, $q_1^*$, at the same wavelength. However, the spectral weight between $\Delta_0$ and $\Delta_{00}$ does not disperse in the ratio map, or in the raw data. The ratio map $q_1^*$ also has the same wavelength as the raw data $q_1^*$. The $q_5^*$ pseudogap peak is enhanced in the ratio map and spread in energy. This enhancement highlights the nematic structure that has been reported previously[25,26].

The individual energy LDOS(*r*) and corresponding FT patterns display the three different spatial modulations (QPI, checkerboard, and pseudogap) and show the transitions between them. Figure 8 shows both the ratio map LDOS(*r*) pattern and the corresponding ratio map FT's. The ratio map serves to highlight the spatial excitations by removing the common LDOS features. A non-ratio map is dominated by the $\Delta_1$ disorder, which masks the spatial excitations, even at low energies. In figure 8 at low energies, in the coherent QPI phase, there is a pattern in the spatial LDOS(*r*) that is dominated by $q_7$, the longest most intense q-vector at low energies in the ratio map. At $\Delta_0$ the majority of ***q***-vectors associated with the octet QPI are gone and $q_1^*$ dominates the map, marking the onset of the checkerboard pattern. In the real space and the ***q***-space data both the energies at $\Delta_0$ and $\Delta_{00}$ show equivalent spatial structure. The intensity of the checkerboard vector drops to half its maximum intensity at $\Delta_{00}$, or -3dB. For $\Delta_1$ the maximum intensity in the peak value of the pseudogap peak $q_5^*$ occurs and marks the third energy scale.

The numerical observables for $\Delta_1$, $\Delta_{00}$ and $\Delta_0$ are plotted in figure 9. The $\Delta_0$ obtained by the LDOS(E) fits matches the $\Delta_0$ where the QPI terminates. The checkerboard -3 dB point and the fitted $\Delta_{00}$ agree quantitatively, as well as $\Delta_1$ and $q_5^*$ peak intensities. The -3 dB point corresponds with a decrease in the checkerboard intensity by a factor of 2. If $\Delta_{00}$ is locally disordered, then this would represent the point where half the field of view has passed their local $\Delta_{00}$, or the average $\Delta_{00}$ of the sample. The checkerboard termination could be caused by the decrease in the intensity of the $q_1^*$ as a function of energy smoothly over the entire field of view, or by $q_1^*$ disappearing in small spatial patches at specific energies that have a distribution in energy. In order to test this the tripartite model will have to be applied to a large number of spectra to determine the disorder in $\Delta_{00}$.

**$(+-\pi/a_0,0) - (0,+-\pi/a_0)$ *k*-space boundary**

QPI studies have shown that the termination of the dispersive QPI pattern occurs at lines connecting $(+-\pi/a_0,0)$ to $(0,+-\pi/a_0)$. In the tripartite model the location of the termination and the resulting non-dispersive section of the spectral density of states is left as a free parameter. The *k*-space points that correspond to the $\Delta_0$ point for both the LDOS(E) and the QPI fits are plotted in figure 10a. The $\Delta_0$ points lie along the line for both the QPI fits and the LDOS(E) fits. The fitting uses a rigid band structure[32] whose chemical potential is obtained through fits to the QPI octet model extracted *k*-space points. These band structure fits, as well as the resulting change in chemical potential with doping, are shown in figure 10b. A change in the band structure parameterization has little effect on the resulting generated LDOS(E). However, it does have a large effect on the $\Delta_0$ *k*-space point that is calculated from the LDOS fit. If there is a slight change in the hopping parameters with doping, then the LDOS $\Delta_0$ *k*-space point can shift, and since fixed hoping parameters with doping are assumed, is the cause of the increase in error with decreasing doping seen in 10a.

The tripartite model assumes that there is a lowest harmonic d-wave gap that exists outside the $(+-\pi/a_0,0) - (0,+-\pi/a_0)$ boundary. This assumption follows from the previous fitting attempts and the assumption that there is a d-wave superconducting gap along the entire Fermi surface. However, there is no evidence from the spatial excitations that such states exist outside the $(+-\pi/a_0,0) - (0,+-\pi/a_0)$ boundary. The pseudogap $q_5^*$ wave vector and $q_1^*$ are both set by the point where the Fermi surface crosses $(+-\pi/a_0,0) - (0,+-\pi/a_0)$ boundary. SI-STM cannot prove that there is no d-wave gap on the Fermi surface outside the boundary and such a signature could be masked by the large amount of high energy $\Delta_1$ gap disorder present at all dopings.

**Discussion:**

The tripartite model reproduces both the LDOS(E) and the QPI data using a common set of parameters and identifies three regions in energy that have unique signatures. The energy scales that mark these three regions are tied to the tripartite models spectral density of states, which can be observed through QPI in *k*-space up to $\Delta_{00}$ in energy and up to the $(+-\pi/a_0,0) - (0,+-\pi/a_0)$ boundary in momentum. In the LDOS(E) these three regions are the low energy d-wave + higher harmonic effective gap, the kink, and the high energy pseudogap peak. In *q*-space, these three are the low energy dispersive QPI octet pattern, the checkerboard modulations (nondispersive $q_1^*$ where $q_1$ stops dispersing) and the pseudogap large scale modulation $q_5^* \sim 3\pi/2a_0$ (nondispersive at the wavelength where $q_5$ disappears)[25]. This large length scales pseudogap pattern has fine structure that locally breaks $C_4$ symmetry

($q_{Bragg}=2\pi/a_0$) of electronic structure within $CuO_2$ unit cells down to $C_2$ symmetry and coexisting with incommensurate conductance modulations locally breaking both rotational and translational symmetries[24–26] ($q_5^* \sim 3\pi/2a_0$).

The three spatial phenomena have signatures in both the ratio map and the raw data, allowing the checkerboard and pseudogap to be separately identified from the dispersive $q_1$ and $q_5$ *q*-vectors. The disappearance of $q_5$ and the change in $q_1$ in the ratio map point to these two *q*-vectors having a different phase from the others.

The gradual nature of the transition between the regions in *q*-space, and the sharp transition in the LDOS(E), point to the possibility of spatially disordered energy transitions. The *q*-space transitions would then be an average over many sharp transitions that occur at different energies depending on the location in the sample. Further individual fitting must be carried out in order to verify this lower energy disorder. The tripartite model will be essential in this study as a robust underlying model is necessary for large scale, accurate, curve fitting of noisy data.

### Other measurements of the gap:

The low energy gap inferred from dispersive coherent quasiparticle scattering by SI-STM differs from that by leading edge extraction by ARPES[34,35], dynamical structure factor extraction by Raman[36] and extrapolated measurements of the nodal slope of the gap structure by thermal conductivity[37]. All three probes spatial resolutions are large compared with the disorder length scale and they report a lowest harmonic d-wave gap in the nodal region. This nodal region is also where ARPES is most reliable and is the only region able to be probed by thermal conductivity measurements. ARPES and Raman also have the ability to measure the gap at the anti-nodes and in some studies they provide evidence that something other than a lowest order d-wave superconducting gap is present. ARPES and Raman, however, must rely on background subtraction techniques[38] and theoretical models[36] to extract both nodal and anti-nodal gap energies. Differences in these techniques between groups are the main cause of the wide range of values and behaviors reported. The most contentious difference between the bulk probes and SI-STM is the presence of the higher order harmonic gap structure at low energies. A possible explanation put forth recently has been the failure of the octet model to accurately extract the underlying *k*-space structure[22]. However, if this were the case then one would expect the LDOS to not be consistent with the QPI as reported here. It is possible that this higher harmonic term is caused by either a tunneling matrix element[39] or a scattering effect[40]. The nature of the ARPES probe makes it insensitive to two particle scattering events[23], since ARPES studies the single particle spectral function, a limitation that SI-STM does not share. The scattering measured by SI-STM is likely the cause of the higher harmonic term[22] present in the SI-STM data.

Recent Raman measurements[36] show that the intensities of the Raman features have qualitative differences that are consistent with coherent quasiparticles being restricted to an arc near the node. They see the percentage of Fermi surface in the arc decrease with decreasing doping. SI-STM measurements[4], as captured by our model, also have a section of the Fermi surface inside the parent compound AF-zone boundary that contains coherent dispersive excitations. These coherent states end at the lines joining $(+-\pi/a_0, 0)$ to $(0, +- \pi/a_0)$ and as the samples are underdoped, the Fermi surface bounded by these lines decreases in length. This matches the qualitative effect that is observed in Raman, although the Raman results show the size of the Fermi surface that contains the coherent quasiparticles to decrease faster with decreasing doping then what is seen in SI-STM. SI-STM, unlike Raman spectroscopy, provides high *k*-space resolution in this region and allows the termination of these collective excitations to be matched to the lines joining $(+-\pi/a_0, 0)$ to $(0, +- \pi/a_0)$ for all dopings under p~0.19.

The Raman measurements[36] of the nodal and anti-nodal gaps that show two energy scales that match the tripartites $\Delta_0$ and $\Delta_{00}$, resolving a smaller nodal gap around $\Delta_0$ and a larger anti-nodal gap at $\Delta_{00}$. The absence of the higher energy pseudogap energy scale as seen in SI-STM gives support to the idea that the pseudogap is a localized state at high energies and would appear in *k*-space, due to localization, at all *k*-vectors. This comparison agrees with the ARPES data[34] that shows a lowest harmonic d-wave superconducting gap over the entirety of the Fermi surface.

Raman also reveals that below p~0.21 that there is a strong two particle interaction[36], an interaction that increases with decreasing doping. While this study contains no data over 0.19 in doping, the shrinking of the kink and disappearance of the checkerboard as doping is increased strongly implies that these effects are not present or weak on the overdoped side of the phase diagram. If the checkerboard modulations represent a two particle scattering phenomena, then it explains why ARPES has difficulty resolving it. Only when the scattering is strong enough does it produce a static structure with large scale (micrometer) phase coherence that ARPES is capable of detecting[41]. Since the checkerboard modulations are extremely strong in $Ca_{2-x}Na_xCuO_2Cl_2$[6], that would explain the checkerboards presence in ARPES data there and its absence from ARPES data in $Bi_2Sr_2CaCu_2O_{8+x}$.

There have been several ARPES studies[35] showing a difference in the dispersion between the node and anti-node. There is also a ARPES report of a lowest harmonic d-wave dispersion[34], which grows with size as doping is decreased and exists over the entire Fermi surface. This study reports a anti-nodal gap energy as a function of doping that is consistent with the measured $\Delta_{00}$ energy values reported here. This implies that the SI-STM data is composed of two separate phenomena. At low energies there is a d-wave gap + a higher harmonic term. When these states reach the lines joining (+-$\pi/a_0$,0) to (0,+- $\pi/a_0$) the quasparticles are scattered strongly establishing a checkerboard modulation and removing them from the superconductive state. This scattering would not be seen by ARPES and would explain why they see a lowest harmonic d-wave gap rather than a d-wave gap terminated by a checkerboard. The identification of the checkerboard modulations as strongly scattered states is also supported by the vortex core studies[28,42,43] where a checkerboard and a kink are formed from enhanced scattering from the vortex. If this is the same phenomena as the checkerboard seen in the underdoped data, it would imply that the vortex core and the checkerboard are regions where the superconductor is scattered strongly by magnetic interactions. One test for this would be to measure the vortex cores wavelength as a function of doping and see if it matches the Fermi surface intersection of (+-$\pi/a_0$,0) to (0,+- $\pi/a_0$).

**Conclusions:**

The tripartite model presented here is not only consistent with other probes of the gap energy, but it also establishes energy boundaries separating three distinct regions and tracks their evolution with doping. Figure 11 summarizes the three energy scales and their dependence on doping. $\Delta_0$ traces out the superconducting dome of coherent Bogoliubov quasiparticles, while $\Delta_1$ follows the pseudogap trajectory. $\Delta_{00}$ is then where the superconducting gap would be without the interference of the parent compound AF-zone boundary, which removes the coherence from the superconductive state by scattering. The real power of our model comes not from its ability to extract structural information from bulk data, but its ability to extract these parameters from individual local spectra. This ability will pave the way for a fully local description of all three electronic structure components and their spatial relationships answering questions about locality of doping, transitions between the phases and their relationship to underlying scattering processes.

**Acknowledgments:** The authors wish to acknowledge the help of JC Davis with comments on the paper, the physics and for allowing us access to the data sets. We would also like to thank Peter Hirschfeld for his helpful discussions, Donald Wilkerson and Ben Hunt for proofreading help and, John Moreland for his support. Dan Dessu has also been most helpful with ARPES discussions and as a trial audience for the work.

**Figure Captions:**

Figure 1: Previous LDOS fitting. Previously LDOS spectra were fit using a classic d-wave function with lifetime broadening. The fits captured the lifetime information presented in the higher energy peak, however at low energies there are departures from the d-wave background. The dashed lines are the classic d-wave with broadening fit and the energy is renormalized to $\Delta_1$ for each curve. The curves are offset for clarity and the black line at zero energy represents the zero differential conductances for each curve. Highest doping is at the top and lowest is at the bottom. The mean $\Delta$ LDOS(E) curves are an average of all the LDOS in a field of view that have the peak at the mean $\Delta$ energy, where $\Delta$ is determined by the position of the peak (not by fitting). In order to calculate these curve the $\Delta$ peak values are determined for a field each data set. All the LDOS(E) data that have the mean $\Delta$ value are averaged together. We refer to this as the $\Delta$ mean LDOS(E). The resolution in energy is set by the energy resolution of the data set and is 2 meV for OD86k, OD89k, and UD74k. The energy resolution is 4 meV for UD45k and UD20k. For OD86k this mean gap represents 5.0% of the field of view. For OD89k = 4.9%, UD74k = 3.6%, UD45k = 7.8%, and UD20k = 5.6%.

Figure 2: The $q$-space information that is collected about the low energy coherent dispersing quasiparticles. Part a) is an example QPI pattern at 20meV from UD74k that shows all 7 QPI vectors labeled in red. Of relevance for this study, are $q_1$ at $\sim\pi/2a_0$ and $q_5$ at $\sim 3\pi/2a_0$. b) previous fits to the QPI data for 5 different dopings. The data is shown as open circles and offset for clarity. The fit to a d-wave gap with a higher harmonic is shown as a solid line. A lowest harmonic d-wave gap is shown for reference as a dotted line. The data does not approach a lowest harmonic d-wave gap even for the over doped samples. In part c) the $k$-space origins of the QPI scattering extracted using the octet model and the ratio map of the data are shown. These $k$-space points show the lack of nodal states in the QPI excitation, the progression of the Fermi surface with change in doping, and the termination of the observed QPI pattern at the $(+-\pi/a_0,0)$ to $(0,+-\pi/a_0)$ line, shown here as a grey dotted line. The last observable points are shown as filled green circles, highlighting the $(+-\pi/a_0,0)$ to $(0,+-\pi/a_0)$ termination.

Figure 3: SI-STM Phenomena and their proposed energy ordering. Across the top examples of the three $q$-space or spatially ordered phenomena observed in cuprates are shown. At low energies the QPI pattern with its familiar 2 fold symmetric 7 vector dispersing pattern exists. At higher energies this is replaced, at least in underdoped samples, with a $q_1^* \sim \pi/2a_0$ pattern known as the checkerboard. At higher energies the $q_5^* \sim 3\pi/2a_0$ pseudogap pattern appears. The bottom panel has both QPI extracted spectral density of states (open circles and left side axis) and the corresponding LDOS(E) (open squares, right axis) at a doping of UD45k. These have been normalized to the gap energy.

Figure 4: The ***k***-space representation of the tripartite model. In order to demonstrate the tripartite model example gap functions are plotted over ¼ of the Brillion zone in a). The higher harmonic gap structure is shown as the solid red low energy arc. This is a clear departure from Cos[2*θ] gap shown as the dotted blue line. Combining these two gaps, there will always be a jump between them, unless the combination happens at the node or anti-node. This leads to the green line that defines a non-dispersive state between the two gaps. This effective gap structure parameterization allows the transition between the two gaps to be varied continuously. This freedom is necessary if one wishes to have a model that can be used to reproduce/fit all of the data. b) An overhead view of a) the red higher harmonic arc ends at $\theta_{cross}$. At this point the non-dispersive jump occurs shown as the green dot. The non-dispersive region ends in higher energy d-wave gap shown in blue. In c), the effective gap for the model as a function of angle projected on a 2D plane. Integrating the tripartite model over the k-space generates the LDOS(E) shown in d), which shows how for the three part gap structure we can generate a kink. In both c) and d) a lowest order d-wave gap and LDOS structure is shown with a dotted line. Both c) and d) are generated using the same parameters.

Figure 5: Individual fits to the tripartite model using separate parameters for QPI and $\Delta$ mean LDOS(E) data for dopings 0.19, 0.17, 0.14, 0.08, 0.06 from bottom to top offset for clarity. a) shows the tripartite model fits to the QPI data. The higher harmonic part is in red, which matches the QPI data. When the QPI terminates at $\Delta_0$ there is a jump shown in green from $\Delta_0$ to $\Delta_{00}$ where the lowest harmonic d-wave gap starts again. On the right the open blue symbols denote the average gap value measured from the LDOS, as well as its FWHM (error bars.) In b) the LDOS fits utilizing the tripartite model show how the model not only fits the low V-shape, but also integrally fits the kink as it grows larger with decreasing doping.

Figure 6: The combined parameterization fits and data. The fits are plotted utilizing the average parameterization of the QPI and the LDOS fitting. Only the positive side of the LDOS is shown. In a), the open circles represent the $\Delta$ mean LDOS for each of our 5 dopings using the same process described in the caption for figure 1. The resulting $\Delta$ mean curves are then normalized to their fitted $\Delta_1$ value. The red represents energies below $\Delta_0$ and is the section of gap that has the higher harmonic contribution. The green is the kink region and is also the energies where the checkerboard is found. Blue represents the high energy pseudogap states. The tripartite model allows us to capture the kink structure, as well as the energies where each observable state (QPI, checkerboard, pseudogap) resides. In b) the QPI plots are shown. The QPI data stops at $\Delta_0$ where the red, higher harmonic gap also terminates. On the right the average LDOS $\Delta_1$ values are plotted as open blue symbols along with their distribution (error bars). In blue, as a dotted line, there is a lowest harmonic d-wave gap for comparison.

Table 1: These are the parameters used in figure 6 for the fits. Here <$\Delta_1$> represents the average of the gap map while $\Delta_{qpi}$ represents the QPI derived anti-nodal gap. Error is shown in parenthesis.

Figure 7: Line cuts through the QPI signal along the atomic direction [01] normalized by total intensity at each energy level. Insert in a) shows the direction of the cut. a) is the raw normalized QPI signal showing both $q_1$, $q_7$ and the checkerboard ($q_1^*$) and pseudogap ($q_5^*$) signatures. Marked with a black dotted line is the $q_1$ dispersion, seen on both positive and negative biases. The black circle marks the appearance of the checkerboard, which corresponds with the kink in the LDOS(E). At negative biases, there is a mirroring of the checkerboard q-vector at higher energies due to setup effects. $q_5$ appears to be also symmetric about zero bias and at positive energies there exists a peak at the same energy as the gap disorder marking the pseudogap. This peak is mirrored on the negative side by the same type of streaking seen at the checkerboard q-vector, also due to the setup effect. b) The ratio map cut along the same direction as a) showing the disappearance of both the large $q_1$ dispersion and the disappearance $q_5$. The checkerboard is much stronger in the ratio map and merges with the shifted $q_1$ below $\Delta_0$. The *q*-vector disorder, near zero at low energies, disappears due to the removal of the setup effects as well as positive and negative energy common feature disorder. The pseudo gap signature also is increased in intensity and breadth of energy.

Figure 8: The real space and *q*-space excitations contained within each of the three zones for the ratio map. At low energies the QPI resides, which has a distinct pattern in both real space (left column) and *q*-space (right column). The yellow circle marks the same place in *q*-space throughout. The low energy termination energy ($\Delta_0$) obtained from LDOS and QPI shows that the quasiparticle interference has faded away, revealing the static checkerboard pattern. The checkerboard pattern remains in place till $\Delta_{00}$, which marks the end of the kink, as well as the -3 dB point of $q_1^*$'s intensity. There is a rise in $q_5^*$ and the localized CuO plane pattern that marks the pseudogap state. The intensity of $q_5^*$ has a peak at $\Delta_1$. As in figure 8, in the ratio map $q_5^*$ has a wide energy tail on either side of its peak.

Figure 9: All measurements of transition energies. a) the two measurements of the pseudogap energy ($\Delta_1$) from the tripartite model fits (table 1) and the peak in $q_5^*$. b) The two measurements of the checkerboard termination ($\Delta_{00}$), the fits and $q_1^*$'s -3dB point. c) coherent QPI termination ($\Delta_0$) measured by the QPI fits and $\Delta_0$ from the LDOS fits.

Figure 10: *k*-space structure: a) termination *k*-space points determined from the fits alone. The top left is the $\Delta_0$ point determined by QPI and the bottom right is the $\Delta_0$ point determined from the LDOS fits. Both show that $\Delta_0$'s *k*-space point lies along the (+-$\pi$/$a_0$,0) to (0,+- $\pi$/$a_0$) line. In b) the *k*-space QPI origins are plotted represented by the open circles. The band structure fits are plotted as solid lines. A ridged band structure was used allowing only the chemical potential to be varied. The chemical potential as a function of doping is shown in the insert.

Figure 11: Evolution of the phases with doping. At low energy there is an excited state represented by red and by QPI. This ends at $\Delta_0$ where checkerboard (green) begins and lasts until $\Delta_{00}$ where the pseudogap state begins shown in blue. $\Delta_1$ here represents the peak of the pseudogap intensity and the average LDOS(E) gap.

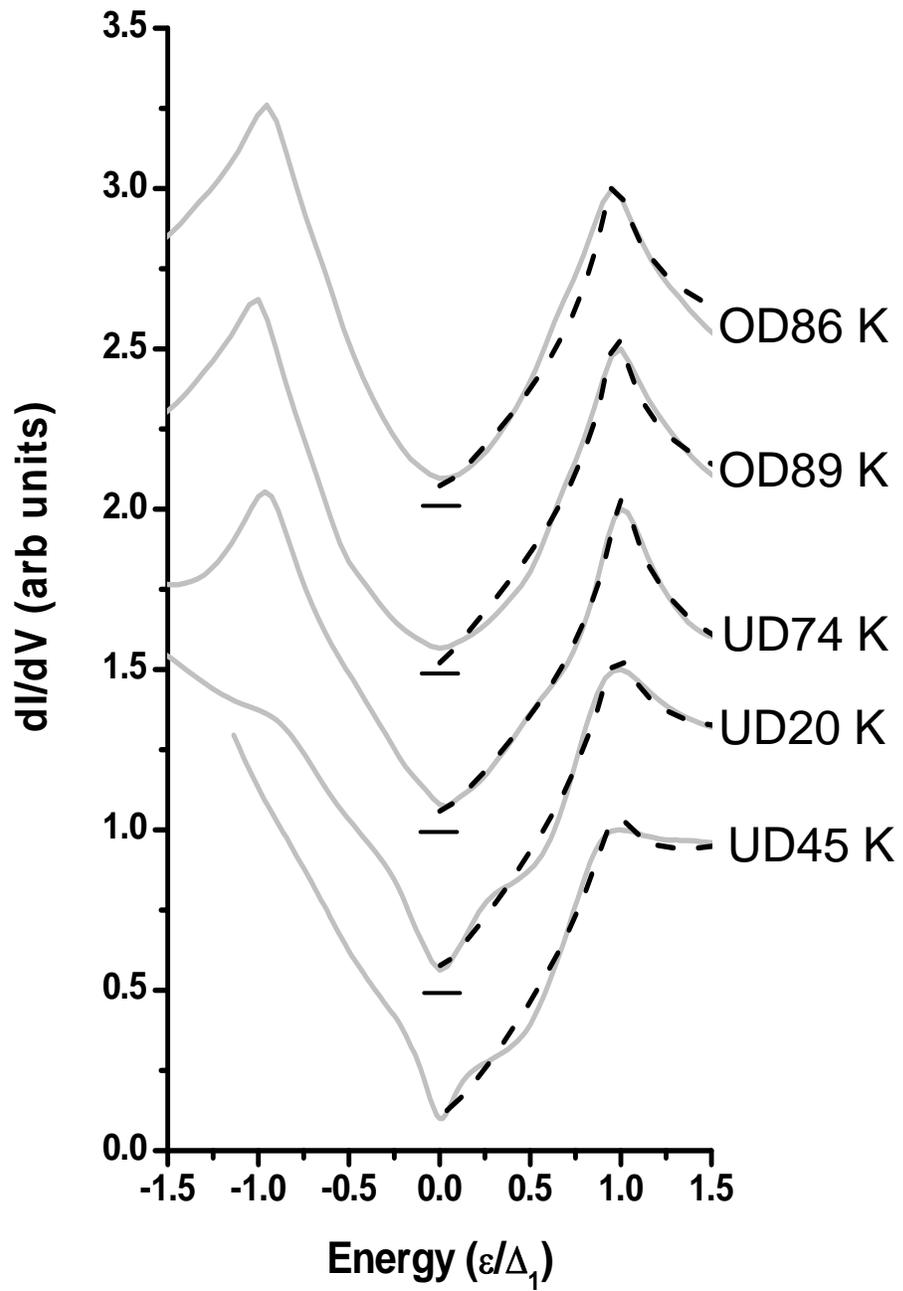

Figure 1

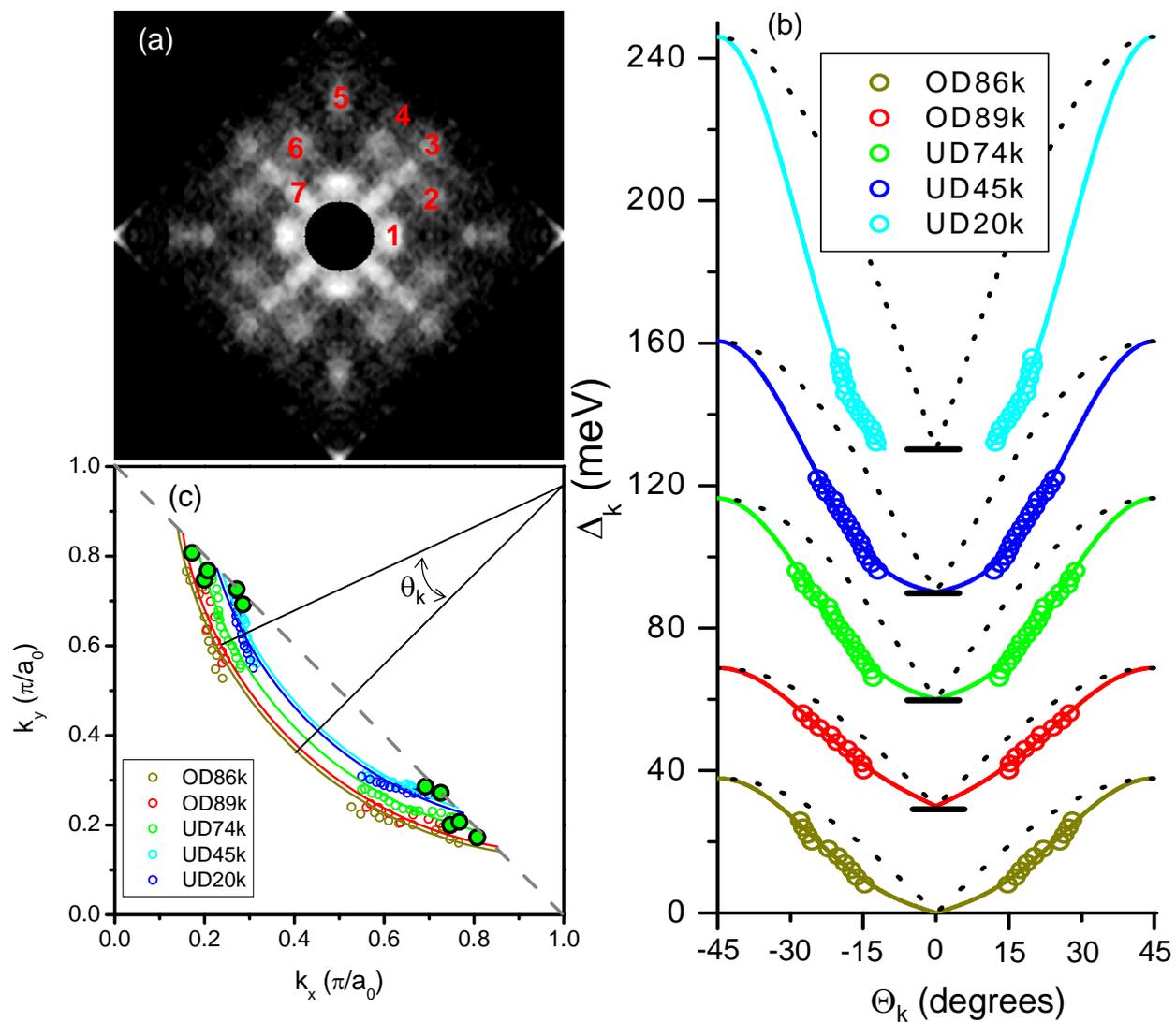

Figure 2

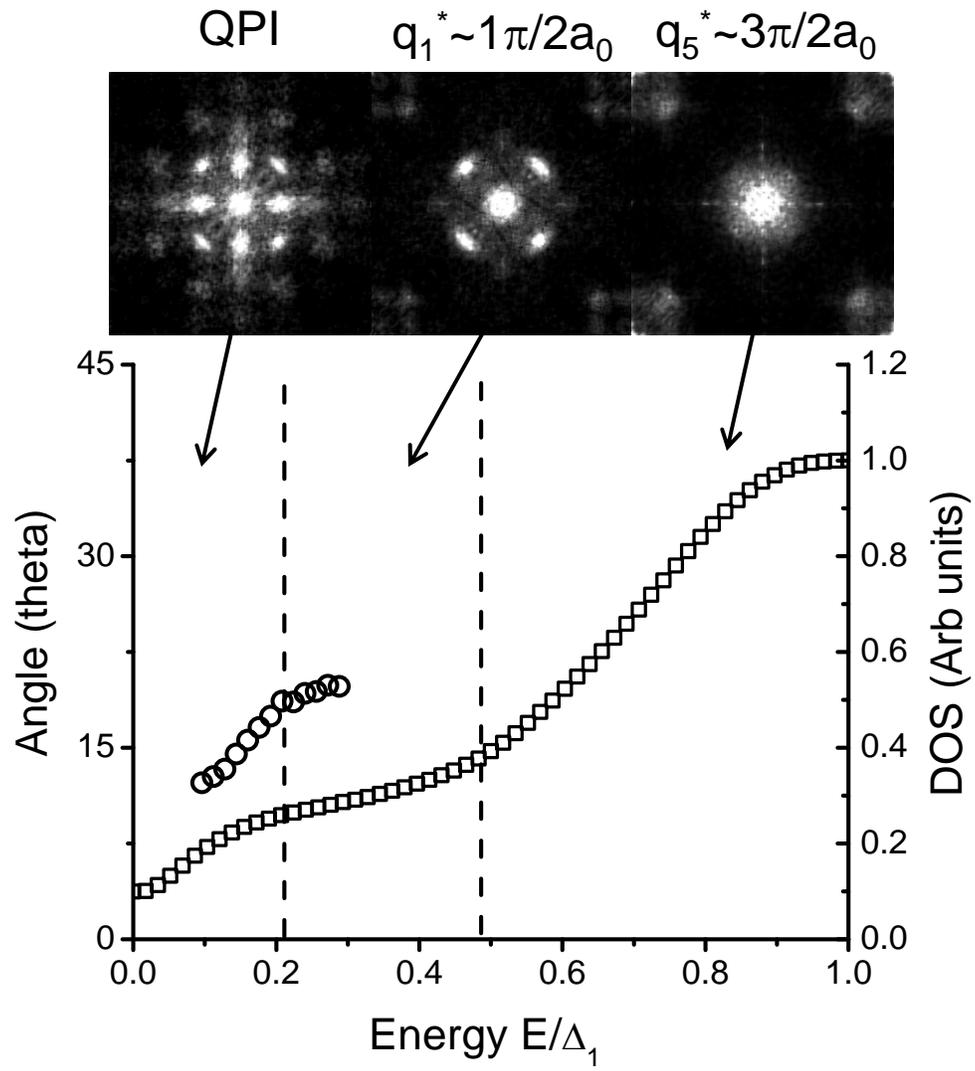

Figure 3

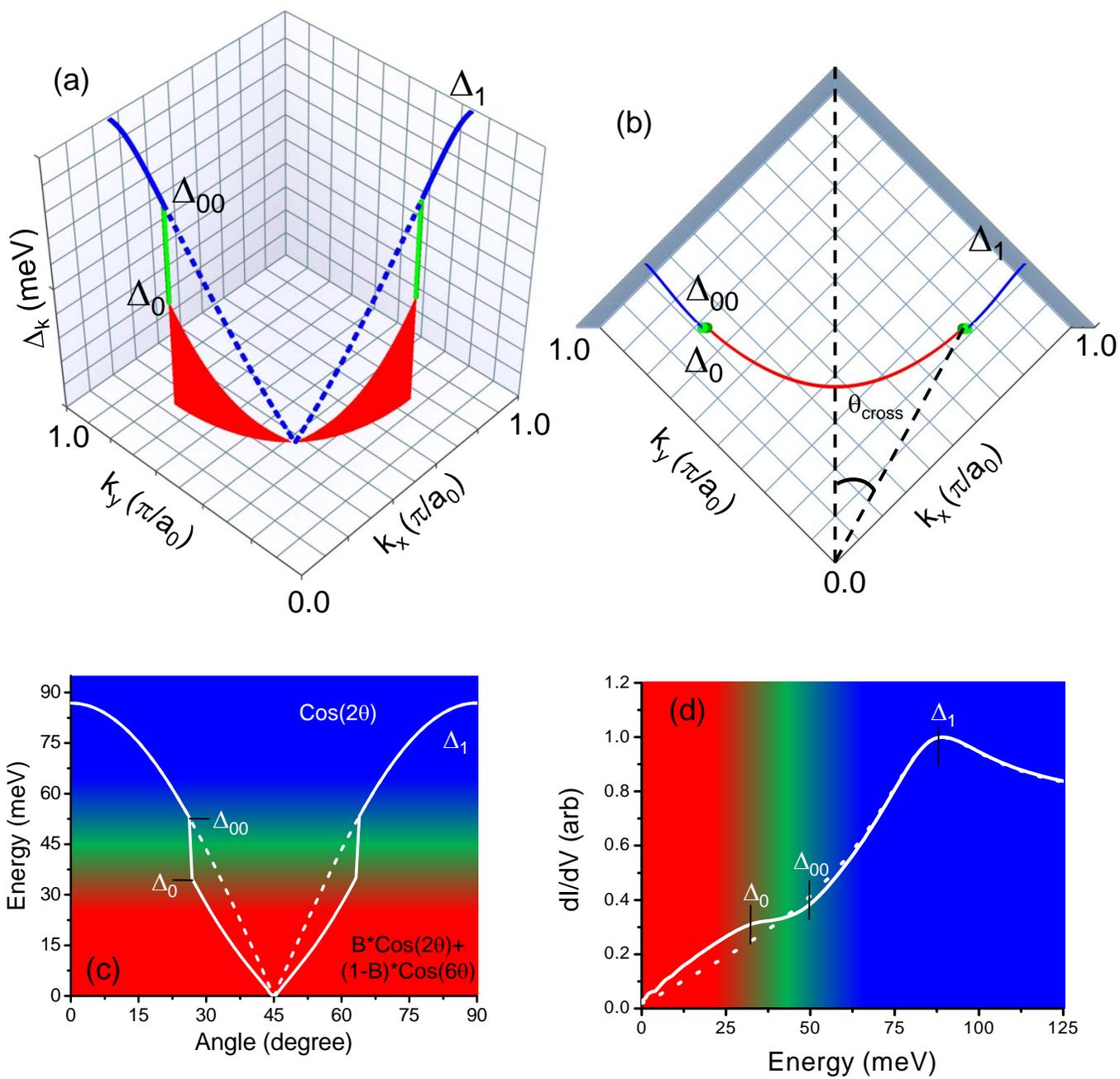

Figure 4

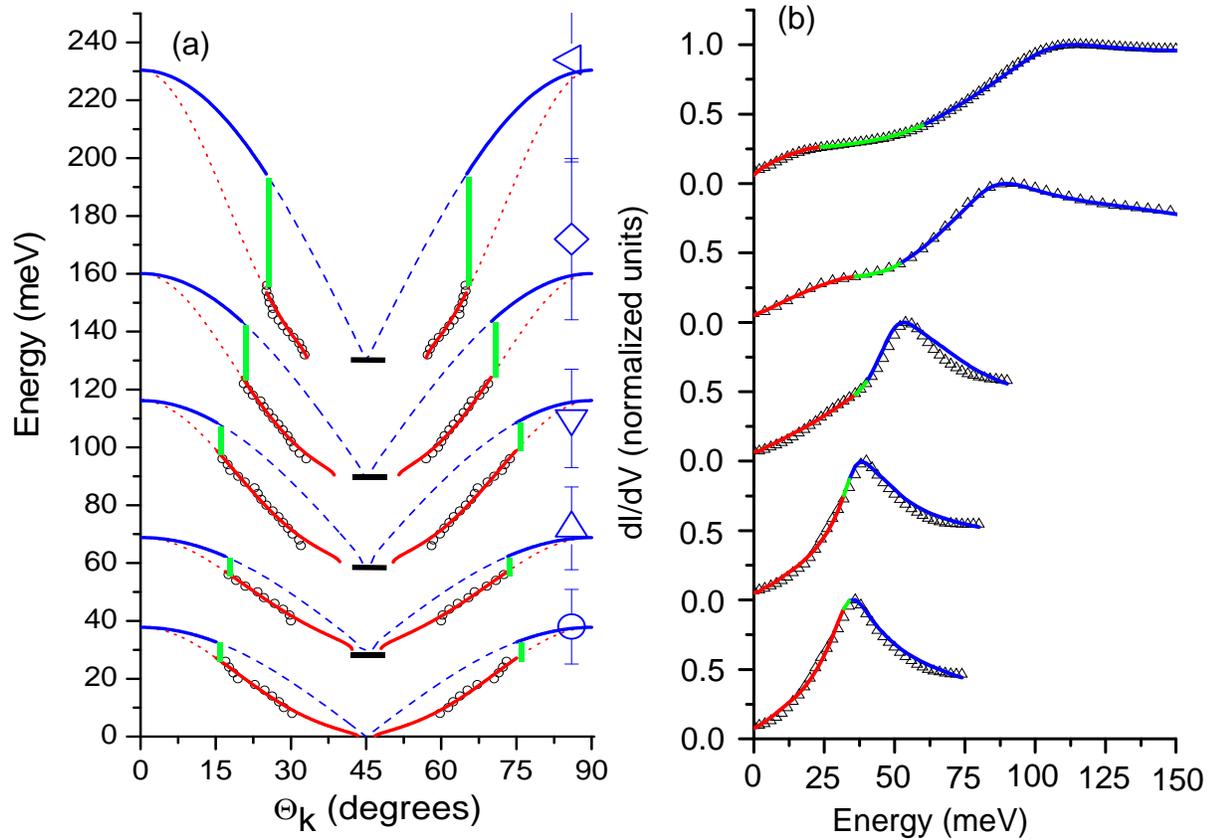

Figure 5

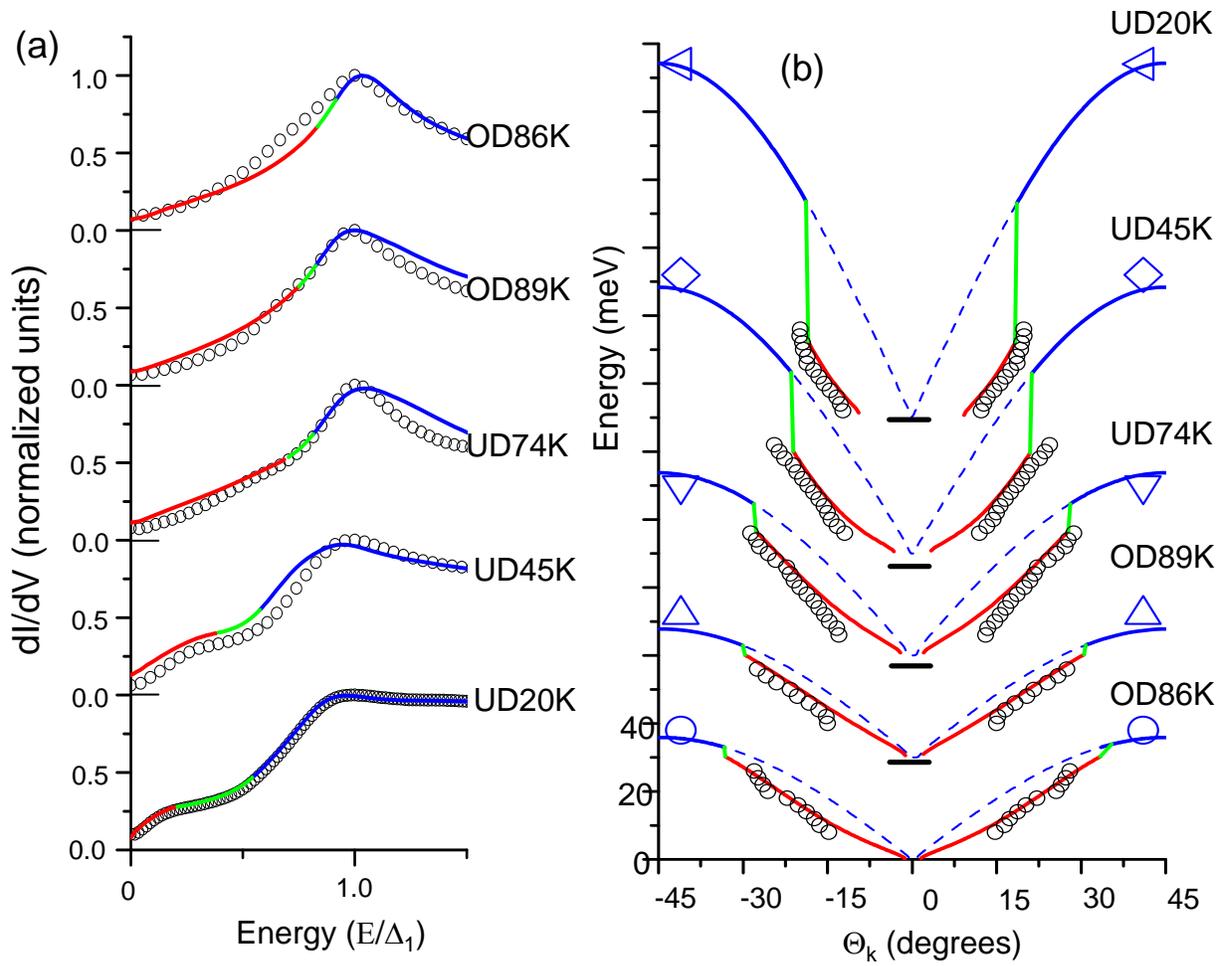

Figure 6

| p | $T_c$ | μ (eV) | $\Delta_1$ (meV) | $\Delta_{qpi}$ (meV) | $\Gamma_1$ (meV) | α | $\Delta_0$ (meV) | B | C/A (meV$^{-1}$) |
|---|---|---|---|---|---|---|---|---|---|
| 0.19 | OD86k | -0.119 | 34.0 | 37.8 | 1.22 (0.12) | 0.125 (0.13) | 30.2 (2.4) | 0.87 (0.035) | -1.62E-5 |
| 0.17 | OD89k | -0.102 | 36.7 | 38.8 | 1.97 (0.84) | 0.11 (0.00) | 30.3 (2.7) | 0.92 (0.045) | -1.70E-5 |
| 0.14 | UD74k | -0.080 | 51.3 | 56.2 | 2.4 (0.08) | 0.12 (0.005) | 36.9 (1.3) | 0.87 (0.055) | -1.55E-5 |
| 0.08 | UD45k | -0.026 | 86.7 | 70.1 | 2.18 (0.52) | 0.14 (0.01) | 35.7 (0.25) | 0.81 (0.075) | 2.67E-6 |
| 0.07 | UD20k | -0.007 | 108.0 | 100.6 | 0.90 (0.13) | 0.15 (0.005) | 22.7 (0.30) | 0.75 (0.02) | 1.26E-6 |

Table 1

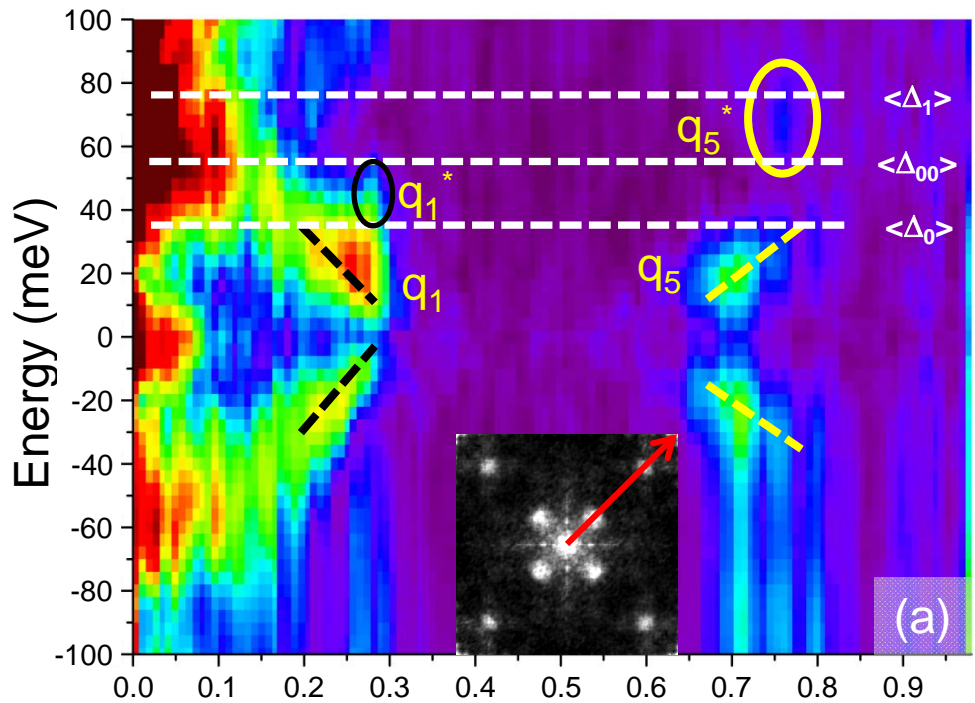
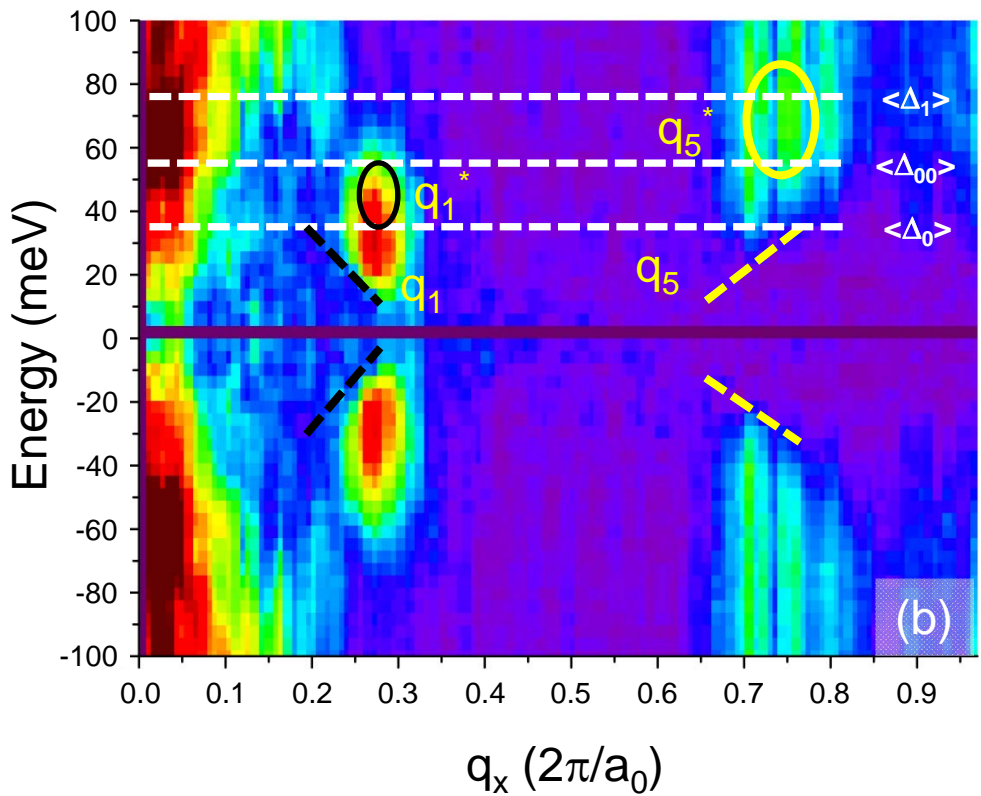
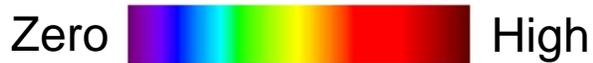

Figure 7

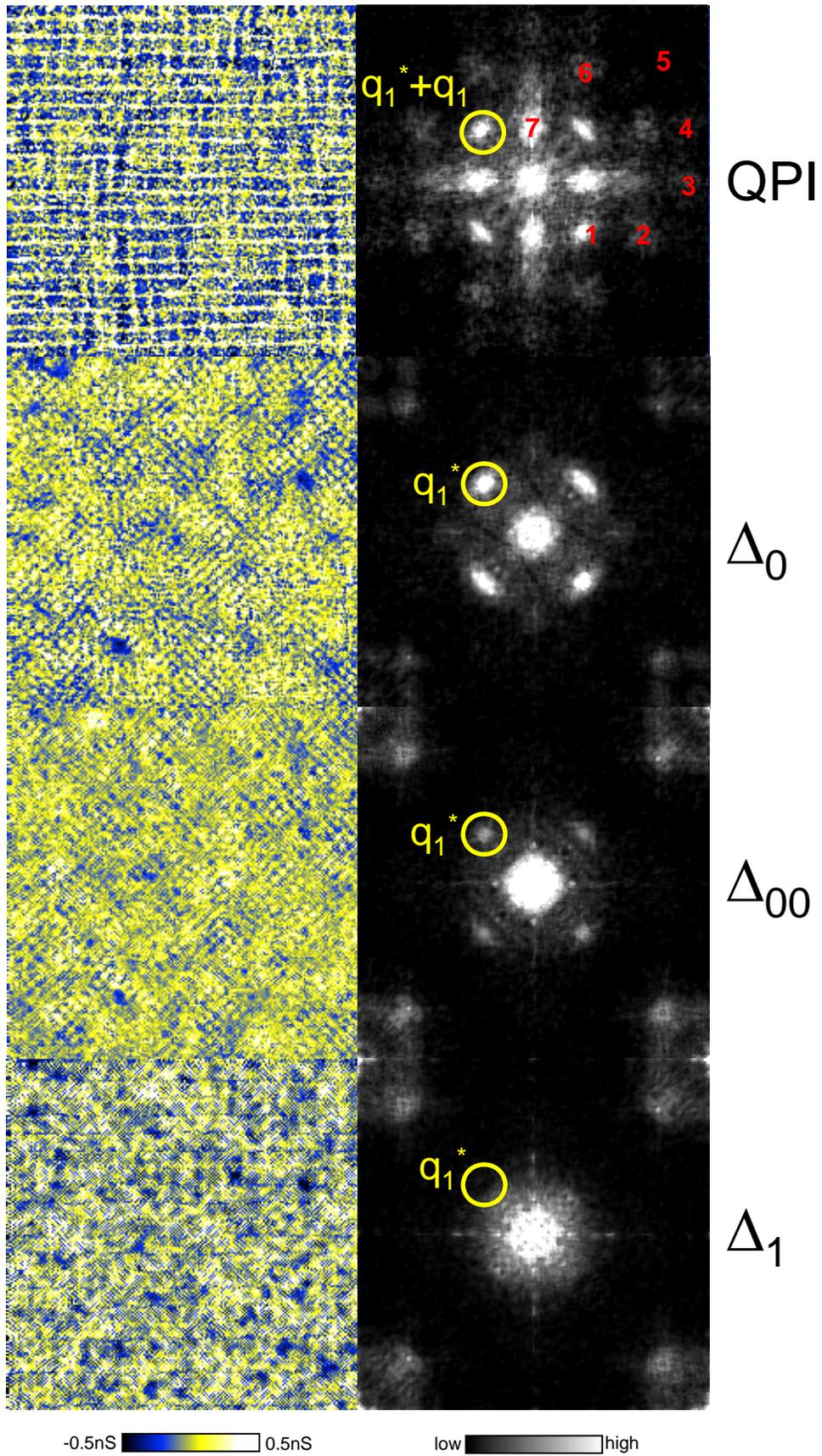

Figure 8

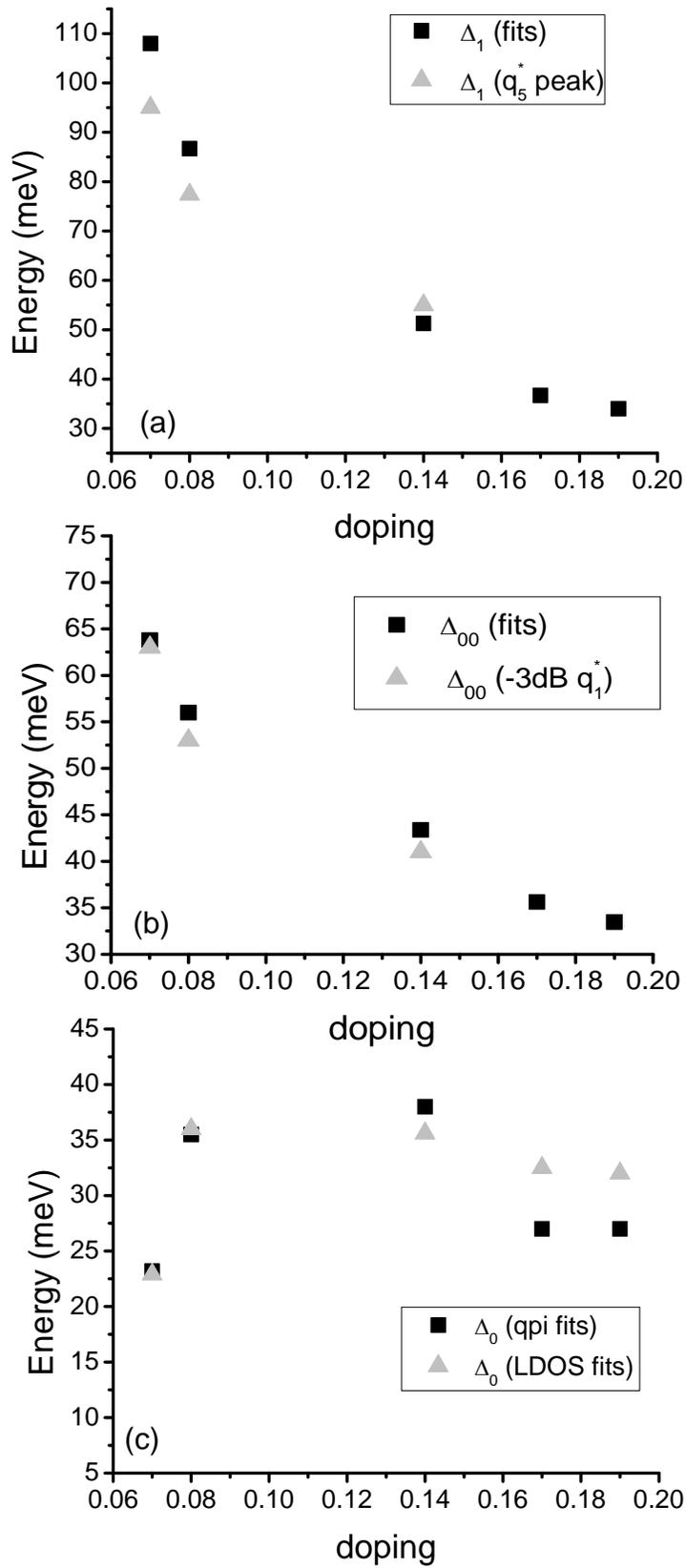

Figure 9

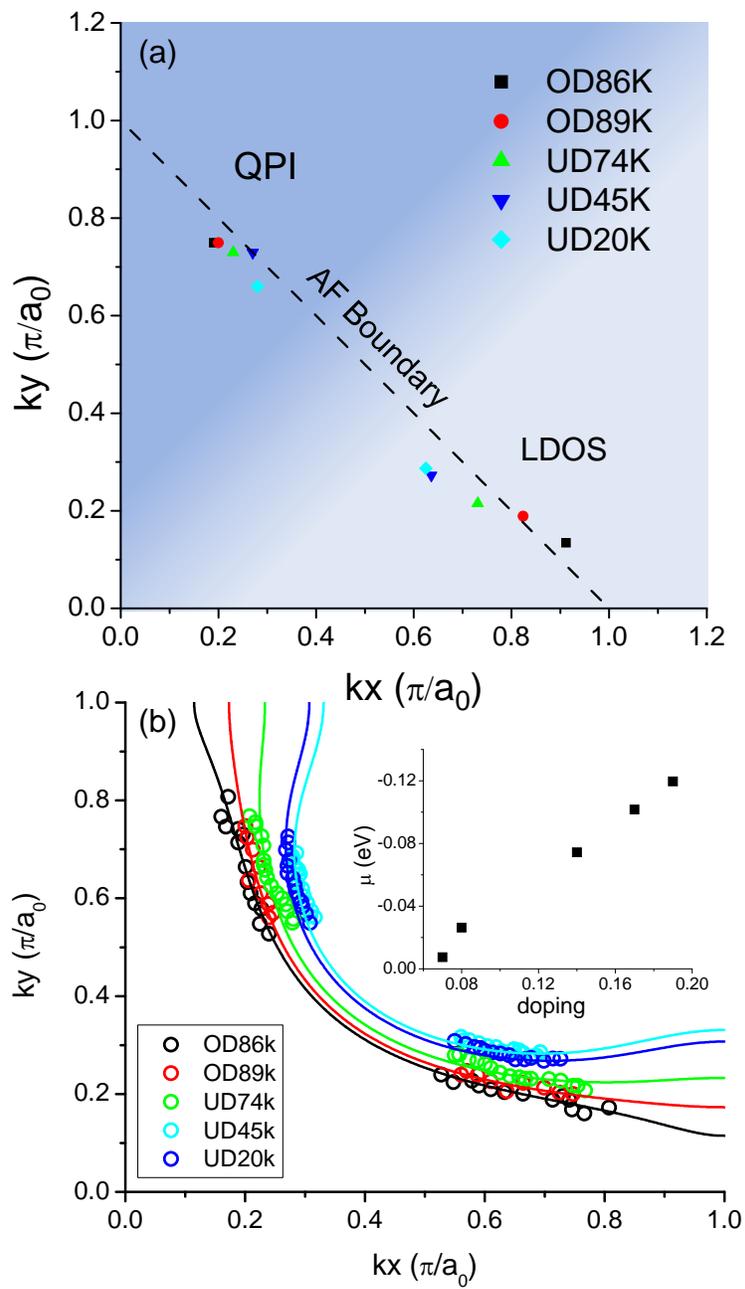

Figure 10

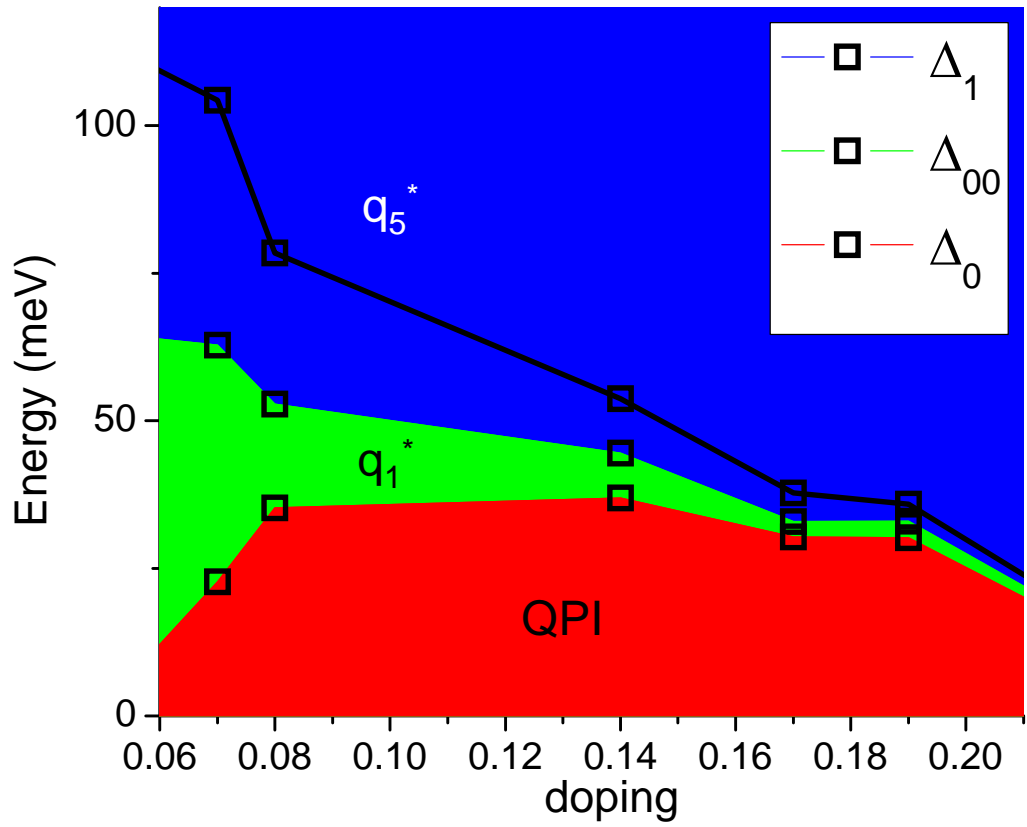

Figure 11